\documentstyle[prl,aps,floats]{revtex}
\input{psfig.tex}

\tighten
\begin{document} 
\draft
\preprint{}
\title{Cosmic concordance and the fine structure constant}
\author{Richard A. Battye{$^{1}$}, Robert Crittenden{$^{1}$} and 
Jochen Weller{$^{2,3}$}}
\address{
${}^1$ Department of Applied Mathematics and Theoretical Physics,
Centre for Mathematical Sciences, University
of Cambridge, \\ Wilberforce Road, Cambridge CB3 OWA, U.K. \\
${}^2$  Department of Physics, University of California,
Davis CA 95616, U.S.A.\\
${}^3$ Theoretical Physics Group, Blackett Laboratory, Imperial
College, Prince Consort Road, 
 London SW7 2BZ,  U.K.
}
\maketitle
\begin{abstract}
Recent measurements of a peak in the angular power spectrum of the
cosmic microwave background appear to suggest that geometry of the
universe is close to being flat. But if other accepted indicators of
cosmological parameters are also correct then the best fit model is
marginally closed, with the peak in the spectrum  at larger
scales than in  a flat universe. Such observations can be reconciled
with a flat universe if the fine structure constant had a lower
value at earlier times, which would delay the 
recombination of electrons and protons and also 
act to suppress secondary oscillations as observed. We discuss
evidence for a few percent increase in the fine structure constant
 between the time of recombination and the present.
\end{abstract}
\date{\today}

\renewcommand{\thefootnote}{\arabic{footnote}}
\setcounter{footnote}{0}

\section{Introduction} 
\label{sect-intro}

Cosmologists have for many years struggled to find a model of the universe 
consistent with all the available evidence.  
Recently, many observations have pointed to the universe being 
spatially flat with cold dark matter (CDM) making up 
approximately a third of the critical density, and the remainder 
dominated by a component with a negative equation of state such as a 
cosmological constant, $\Lambda$.  
This confluence of evidence, which includes measurements of the expansion  
and acceleration of the universe, bounds on the age of the universe, and 
constraints from large scale structure and galaxy clusters, has been called `cosmic concordance'
\cite{concord}.

The restriction to spatially flat models was originally motivated 
by theoretical arguments, in particular to be consistent with 
inflationary models of the early universe. 
However, measurements of the anisotropy in the 
cosmic microwave background (CMB) have given strong 
observational support to this assumption. 
The position of the first Doppler peak in the CMB power spectrum 
is sensitive to the spatial geometry  
and the epoch of last scattering of the CMB photons relative to the present 
age of the universe\cite{cmb-theory}.
Recent measurements by the balloon 
borne detectors BOOMERanG\cite{boom1} (B98) and MAXIMA\cite{maxima1} (M99)
indicate an increase in power at  
angular scales of approximately two degrees, very close to the expected 
position of the first Doppler peak in spatially flat models. 

While providing tentative confirmation of a flat universe, the
new data has also raised many questions as to the precise viability of
the concordance\cite{HU,WSP,TZ,boom2}. In
particular, the position of the first peak  as detected by BOOMERanG
($\ell_{\rm peak}=197\pm 6$)
appears to be at slightly larger scales than expected in generic
flat models --- a conclusion which is
only slightly weakened by the inclusion the less sensitive MAXIMA
data~\cite{maxima2}.  In addition, both data sets indicate  secondary
Doppler peaks that are much less pronounced than expected in models
compatible with primordial Big Bang
Nucleosynthesis (BBN)~\cite{BBN,BT}. There are, of course, numerous possible
explanations for these discrepancies. For example, the shift of the
peak to larger scales could indicate a universe which is slightly
closed~\cite{WSP,boom2}, while the suppression of the second peak
could be evidence that the baryon density is higher than has been
indicated by BBN\cite{TZ,boom2}.  However, these solutions require
either giving up the elegance of spatially  flat models or run into
direct conflict with other cosmological measurements, particularly
those related to BBN. 

Another possible solution which addresses both unexpected features of the data 
is to delay the epoch of last scattering.  This would increase the size of
the sound horizon at last scattering and shift the first peak to
larger scales while keeping a spatially  flat universe.  It would also
simultaneously increase the density of baryons relative to that of
the photons during the epoch of last scattering, suppressing the
amplitude of the second peak.  Within the standard framework, it is 
rather difficult to change the time of decoupling since it would require 
a mechanism, astrophysical~\cite{PSH} or
otherwise~\cite{Sciama:82,Salati:84,WBA}, which could delay the
formation of neutral hydrogen. Peebles {\it et al}. \cite{PSH} 
suggested that non-linear structures at extremely high redshift could
act as a source for Lyman-$\alpha$ photons which photo-ionize the hydrogen. 
However, this is very unlikely in standard adiabatic models for
structure formation, though it is a possibility if the initial fluctuations 
were non-Gaussian. 

In this paper, we focus on another possible mechanism for delaying
photon decoupling:  that the electrons and protons might have been more
weakly bound at high redshifts than they are today.   In particular,
we consider whether the observations contain evidence for a running of
the fine structure constant, $\alpha=e^2/(4\pi\hbar c)$,
between the time of recombination, where the CMB last scattered, and
the present epoch. Changes in $\alpha$ modify the parameters governing
recombination~\cite{alpha1,alpha2}, which, depending on the sign, can lead to
early ($\delta\alpha(t_{\rm rec})=\Delta\alpha(t_{\rm rec})
/\alpha>0$) or delayed
($\delta\alpha(t_{\rm rec})<0$) recombination. Here, we have defined
$\Delta(t)=\alpha(t)-\alpha(t_0)$, where $t$ is cosmic time and we
denote $t_0$, $t_{\rm rec}$ and $t_{\rm nuc}$ throughout as the
times of the present day, recombination and nucleosynthesis respectively. 

Such variation of physical constants has been the subject of much 
attention, both observational and theoretical, in recent years. The
theoretical motivation comes from String or M-theory in models where
there are compact extra dimensions. These extra dimensions may have
either stabilized before recombination ($\delta\alpha(t_{\rm rec})=0$)
or they may still be rolling down their potential, 
causing all the coupling constants to vary ($\delta\alpha(t_{\rm
rec})\ne 0$). The
standard way in which to do this is using a scalar field known as the
dilaton, but no stabilizing potential has ever been derived from
anything which could be described as a candidate fundamental theory;
all proposed stabilizing mechanisms appear to be {\it ad hoc}. 
Slow variation in $\alpha$ could, therefore, be considered at some
level as a prediction of fundamental theory. It has also been
pointed out~\cite{Beck,BM} that models in which $\alpha$ varies can  be
thought of as models with a varying speed of light~\cite{AM,AB}.

Of course, limits exist on changes in $\alpha$ due to various
terrestrial, astrophysical and cosmological arguments. Terrestrial
limits come primarily from elements which have long-lived $\beta$
decay,  atomic clocks and the OKLO natural nuclear
reactor~\cite{oklo}, for which the limit is $-0.9\times
10^{-7}<\delta\alpha<1.2\times 10^{-7}$ over a time period of around
1.8 billion years\footnote{The astrophysical and cosmological limits
that we shall discuss correspond to a limit on $\delta\alpha$ over a
particular redshift range. If considered in this way the OKLO
constraint can be thought of as being at $z\approx 0.1$}, although
this is model and theory dependent since the limits are sensitive to
possible simultaneous variations in other coupling constants. Cosmological
limits come from the Helium abundance in BBN and quoted limits can be
expressed roughly as $|\delta\alpha(t_{\rm
nuc})|<10^{-2}-10^{-4}$~\cite{KPW,Barrow,Berg}
at $z\approx 10^{9}-10^{10}$, although this is again highly model dependent;
we shall return this issue in a detailed discussion below. 
Astrophysical limits arise from systems which absorb quasar emissions
over a wide redshift
range of $z\sim 0.1-3$~\cite{obs}. After many years of deriving upper bounds on
$\delta\alpha$, a statistical detection of $\delta\alpha=(-1.1\pm
0.4)\times 10^{-5}$ has been claimed recently~\cite{webb} due to measurements
of relativistic fine structure in absorption systems in the range
$0.6<z<1.6$, with more data soon to be published.

In what follows we will work within the framework of spatially flat
$(\Omega_{\rm tot}=1)$ $\Lambda$CDM models with matter density
(in units of the critical density $\rho_{\rm crit}=3H_0^2/8\pi G$)
given by $\Omega_{\rm m}$ and cosmological constant density
$\Omega_{\Lambda}=1-\Omega_{\rm m}$. Similarly, the baryon density is
denoted $\Omega_{\rm b}$ and the rest of the matter is assumed to be 
dominated by
CDM, with no hot dark matter (HDM) component, $\Omega_{\nu}=0$. The Hubble
constant will be  parameterized by $H_0=100h{\rm km}\,{\rm
sec}^{-1}\,{\rm Mpc}^{-1}$. We shall assume that the initial
scalar fluctuations that are measured were created during an epoch of cosmic
inflation, and that they are almost scale invariant with spectral
index, $n_{\rm S}$ and amplitude $A_{\rm S}$. At this stage we will
ignore the possibility of a tensor component to the fluctuations and
also the possibility of early reionization, and for preciseness we
shall assume that the temperature of the CMB is $T_{\rm cmb}=2.726K$,
the fraction of primordial $^{4}{\rm He}$ is $Y_{^{4}{\rm He}}=0.24$ and the
number relativistic degrees of freedom is $N_{\nu}=3.04$.  For rest of
the paper we denote $\delta\alpha(t_{\rm rec})=\delta\alpha$, unless
otherwise specified.

\section{Varying $\alpha$ and the CMB}
\label{sect-cmb}

\begin{figure}
\centerline{\psfig{file=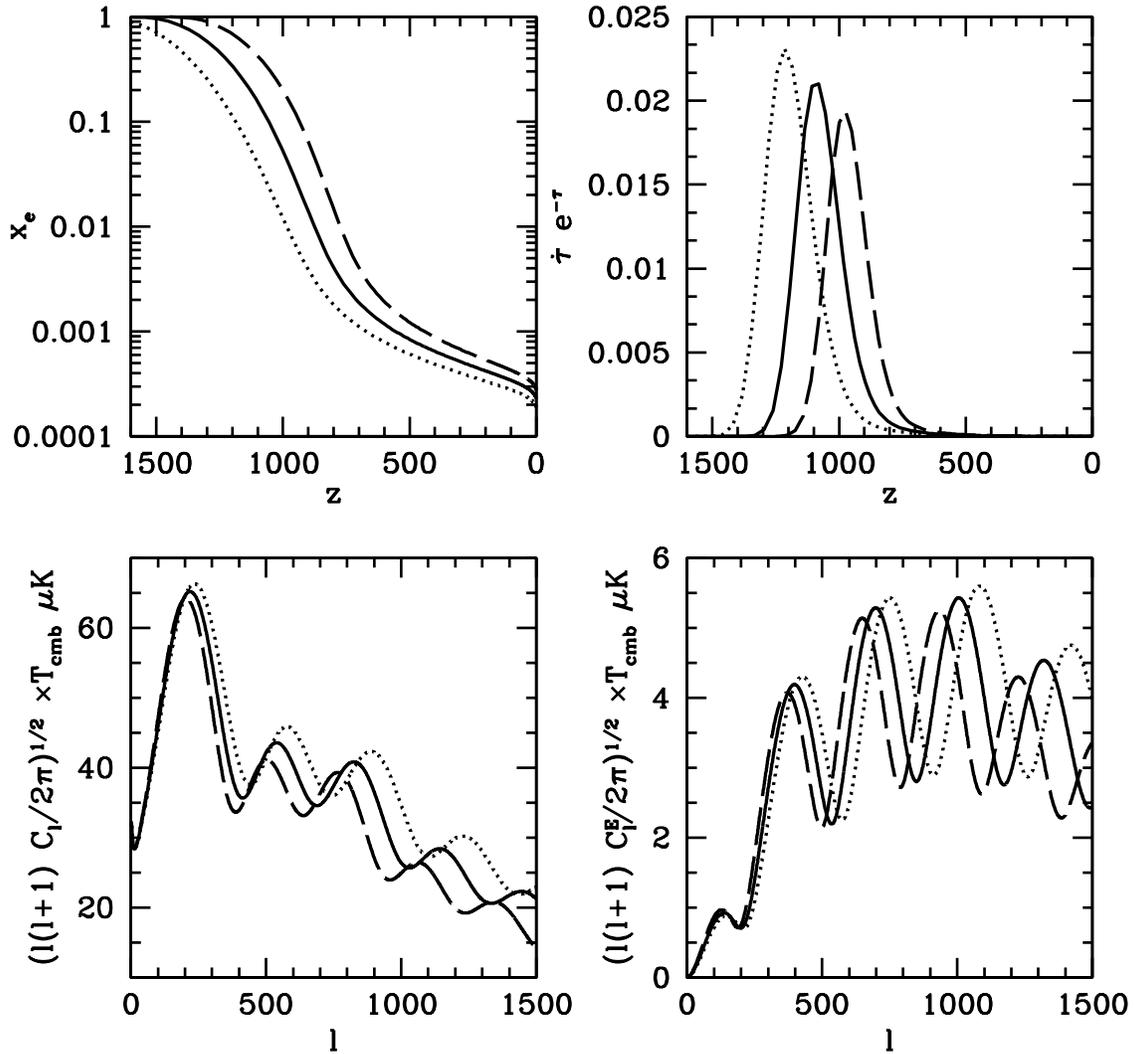,height=6.0in,width=6.0in}}
\caption{The effects of varying $\delta\alpha$ on a standard
$\Lambda$CDM type model as described in the text. In the top-left is
the ionization fraction $x_{\rm e}$ and  in the top-right is the
visibility function $\dot\tau\exp[-\tau]$, both specified as functions
of redshift, $z$. The bottom two curves are
the angular power spectrum of CMB temperature anisotropies 
on the left and that of the
polarization on the right. The solid line corresponds to
$\delta\alpha=0$, the dashed line to $\delta\alpha=-0.05$ and
the dotted line to $\delta\alpha=0.05$.  Note that if
$\delta\alpha<0$ then the peaks are shifted to smaller $\ell$ and 
the amplitude of the second peak is suppressed relative to the first.}    
\label{fig-example}
\end{figure}

To make a quantitative analysis of the effects of changing $\alpha$ on the
CMB anisotropies, we to modify the linear Einstein-Boltzmann solver
CMBFAST~\cite{cmbfast}. 
Here we follow the treatments outlined in
Hannestad~\cite{alpha1} and Kaplinhat
{\it et al}~\cite{alpha2} and confirm their results. 
Changing $\alpha$ modifies the strength of the
electromagnetic interaction and therefore the only effect 
on the creation of CMB anisotropies is via the modifications to the
differential optical depth of photons due to Thomson scattering,
\begin{equation}
\dot{\tau} = x_{\rm e}n_{\rm e}c \sigma_{\rm T}\,,
\end{equation}
where ionization fraction, $x_{\rm e}$, is the fraction of the number density
of free electrons to their overall number density $n_{\rm
e}$, and $\sigma_{\rm T}$ is the Thomson scattering
cross-section. The
ionization fraction, $x_{\rm e}$, is dependent on the temperature of the
electrons, $T_{\rm e}$, and therefore on the expansion rate of the
universe $a(t)$.
Modifying $\alpha$ has a direct effect on the optical depth via
the Thomson scattering cross section, 
$\sigma_{\rm T} = 8\pi \alpha^2 \hbar^2/(3m_{\rm e}^2c^2$). 
It also indirectly effects $\dot{\tau}$ by modifying the
temperature dependence of 
$x_{\rm e}$. These two effects change $T_*$, the
temperature at which last scattering takes place, and  $x_{\rm e}(t_0)$, the
residual ionization that remains after recombination, both of which
influence the CMB anisotropies.

The change of the ionization fraction with $\alpha$ results from modifying 
the interaction between electrons and protons.  
The net recombination rate
of protons and electrons into hydrogen is given by \cite{Peebles:68}
\begin{equation}
t_{\rm rec}^{-1} = \alpha_{\rm e}x_{\rm e}n_{\rm e}C - \beta_{\rm
e}\left(1-x_{\rm e}\right)e^{-3B_1/4T_{\rm e}}C\,.
\label{recrate}
\end{equation}
where $B_1$ is the binding energy of the hydrogen
ground state given by  $B_1= \alpha^2m_{\rm e}c^2/2 (\approx 13.6 {\rm eV}$
for $\alpha=\alpha(t_0)$), and $\alpha_{\rm e}$, $\beta_{\rm e}$, $C$
are constants quantifying recombination, ionization and the two photon
decay in to the ground state.
This is the net rate for recombination to and
ionization from {\em all} states of the hydrogen atom. Recombination
to the ground state can be neglected since such a process
immediately creates a Lyman-$\alpha$ photon which reionizes another
hydrogen atom, although one has to take into account Lyman-$\alpha$ photons
which are redshifted out of the resonance line and also that the ground
state can be reached by two photon decay~\cite{Peebles:68}. The
recombination rate to all other excited levels is~\cite{Gould:70}
\begin{equation}
\alpha_{\rm e} = 2A\left(\frac{2T_{\rm e}}{\pi m_{\rm
e}}\right)^{1/2}\frac{B_1}{T_{\rm e}}
\phi^{\prime}\left(\frac{T_{\rm e}}{B_1}\right)\bar{g}\, ,
\end{equation}
where $A = 2^5 3^{-3/2}\alpha^3 \pi A_0^2 = 2.105\times 10^{-22}\,{\rm
cm}^2$, with the Bohr radius $A_0 = \hbar / \alpha m_{\rm e}c =
0.529\, {\rm \AA}$ and $m_{\rm e} = 0.511\,{\rm MeV}$ the electron
mass. The Gaunt factor $\bar{g}\approx0.943$ is due to quantum corrections
of the radiative process and is only weakly dependent on $\alpha$. As
in refs.~\cite{alpha1,alpha2} we have ignored the effects of in change in
$\alpha$ on this correction. 
The function $\phi^\prime (t_{\rm e})= 0.5(1.735- \ln
t_{\rm e}+t_{\rm e}/6)-\exp(1/t_{\rm e})E_1(1/t_{\rm e})/t_{\rm e}$
comes from summing up the interaction cross sections from all excited
levels \cite{Gould:70}. $E_1$ is the exponential integral function and
$t_{\rm e} = T_{\rm e}/B_1$. The ionization rate is related to the
recombination rate by detailed balance
\begin{equation}
\beta_{\rm e} = \alpha_{\rm e}\left(\frac{m_{\rm e} T_{\rm e}}{2
\pi}\right)^{3/2}e^{-B_2/T_{\rm e}}\, ,
\end{equation}
with $B_2 = B_1/4$ the energy of the lowest lying excited, $n=2$,
state. The correction due to the redshift of Lyman-$\alpha$ photons
and the two photon decay is given by $C=(1+K{\cal D}n_{1s})/(1+K{\cal
D}n_{1s} + K\beta_{\rm e}n_{1s})$, with $K =
\lambda_\alpha^3a/(8\pi{\dot a})$, $\lambda_\alpha = 16\pi\hbar/(3m_{\rm
e}\alpha^2c)$ the wavelength of the Lyman-$\alpha$ photons, ${\cal D} =
8.23 {\rm s}^{-1}$ the net rate of the two photon decay with ${\cal D}
\propto \alpha^8$ \cite{alpha1,alpha2} and $n_{1s} = (1-x_{\rm e})n_{\rm e}$
the number density of atoms in the $1s$ state. 

We have incorporated these dependencies on $\alpha$ into CMBFAST and 
the results are illustrated for a simple $\Lambda$CDM model with
$\Omega_{\rm m}=0.3$, $h=0.65$, $\Omega_{\rm b}h^2=0.019$ and $n_{\rm S}=1$
in Fig.~\ref{fig-example}. On examination of the curve for $x_{\rm
e}(z)$, we see that if the $\alpha$
is increasing with time ($\delta\alpha<0$) then the epoch of
recombination is delayed, whereas if it is decreasing ($\delta\alpha>0)$ 
then recombination happens much earlier. The visibility
function $\dot\tau\exp[-\tau]$ quantifies the probability that a given
photon observed today was last scattered at the specified
redshift. Hence, one could loosely define the epoch of last scattering
to be the maximum of the visibility function. By this definition, the
$\pm 5\%$ shifts in $\alpha$ illustrated in Fig.~\ref{fig-example}
correspond to shifts in the epoch of last scattering by about $\mp 100$ in
$z$, from the value of $z_{\rm rec}\approx 1100$ for
$\delta\alpha=0$. We also studied the effects of changing 
$\alpha$ in the process of Helium recombination and found that they
are negligible as long as $\alpha$ varies in the range we discuss in
this paper.

The shift in the epoch of recombination has a number of implications
for the spectrum of CMB anisotropies~\cite{HSSW}.  First, the angular
positions of the primary and  subsequent peaks in the spectrum are
determined by the physical scale  of the sound horizon for photons at
the time of last scattering. In particular, the position of the first
peak is given by ${\ell}_{\rm peak}\approx \pi\eta_0/(c_{\rm
s}\eta_*)$, where $\eta_0$ is the conformal time of the present day,
$\eta_*$ is that of last scattering and $c_{\rm s}$ is the sound speed
of the photon-baryon fluid around $\eta_*$. In a model where
$\delta\alpha(t_{\rm rec})< 0$, $\eta_*$ is increased while $c_{\rm s}$ is
reduced by a smaller amount due to the larger fraction of baryons at
last scattering, and $\eta_0$ does not change.  Hence, the first peak
in the CMB anisotropies is moved to larger scales, or smaller
$\ell_{\rm peak}$.  Similarly, if $\delta\alpha>0$, $\eta_*$ and
$c_{\rm s}$ are affected in the opposite way and the peak moves to
smaller scales (larger $\ell_{\rm peak}$.)  A reduction of $l_{\rm
peak}$ from 240 to 200. as suggested by the B98 data, could be
achieved by a $16\%$ increase in $c_{\rm s}\eta_*$. Such effects would
be degenerate in parameter space with modifications to
$\Omega_{\rm tot}$, which we have ignored.

Other effects of this shift are changes in the modulation of the peak
heights by baryon drag~\cite{HS}, to the photon diffusion damping
length~\cite{silk}, and to the time between matter domination and last
scattering, which lead to subtle degeneracies between $\delta\alpha$
and $\Omega_{\rm b}h^2$ or $\Omega_{\rm m}h^2$. 
The modulation of the peaks heights is determined by the relative
density of baryons to photons at $\eta_*$,   $R_*=3\rho_{\rm
b}/(4\rho_{\gamma})\propto a_*\propto T_*^{-1}$,
where $T_*$, the temperature at which recombination
takes place, is roughly proportional to the binding energy of the
electrons $T_* \propto B_1 \propto (1+\delta\alpha)^2$. 
Hence, one might think that the effects of increasing the baryon  density,
$\Omega_{\rm b}h^2$, can be accomplished by decreasing  $\alpha^2$ by
the same amount.  However, reducing $\alpha$ and delaying
recombination also results in an increased diffusion photon length,
which also could be caused by  a increase in $\Omega_{\rm b}h^2$. The
degeneracy between $\delta\alpha$ and $\Omega_{\rm b}h^2$ is,
therefore, a complicated one and is likely depend on the scales probed
experimentally.  Finally,
delaying the time of recombination will alter the ratio of matter to
radiation when the photons are last scattered, and so will have effects
similar to changing $\Omega_{\rm m}h^2$.  

\section{Integrated Probability Distributions} 
\label{sect-int}

We are now in a position to calculate likelihood functions given the
CMB data  for $\delta\alpha\ne 0$. 
In models where $\alpha$ is constant, the B98 data, taken on their own, 
prefer spatially closed models with $\Omega_{\rm tot} \approx 1.3$
\cite{boom2}.  In addition, many other parameters take similarly
questionable values (e.g., $t_0\approx 7.6{\rm Gyr}$) when the CMB data  
is not supplemented with other priors.  
For example, measurements of the cosmological distance ladder 
indicate that the Hubble constant is roughly between $h=0.45-0.85$
(which we will take to be the $95\%$ confidence range,) 
while the observed light element 
abundances and BBN indicate the baryon density is 
$\Omega_{\rm b}h^2 = 0.019 \pm 0.0024$ ($95\%$ conf. level)
~\cite{BBN,BT}.
If we restrict to models that are spatially flat,
the CMB data prefer values of these parameters  
in excess of that found by the direct measurements.

\begin{figure}
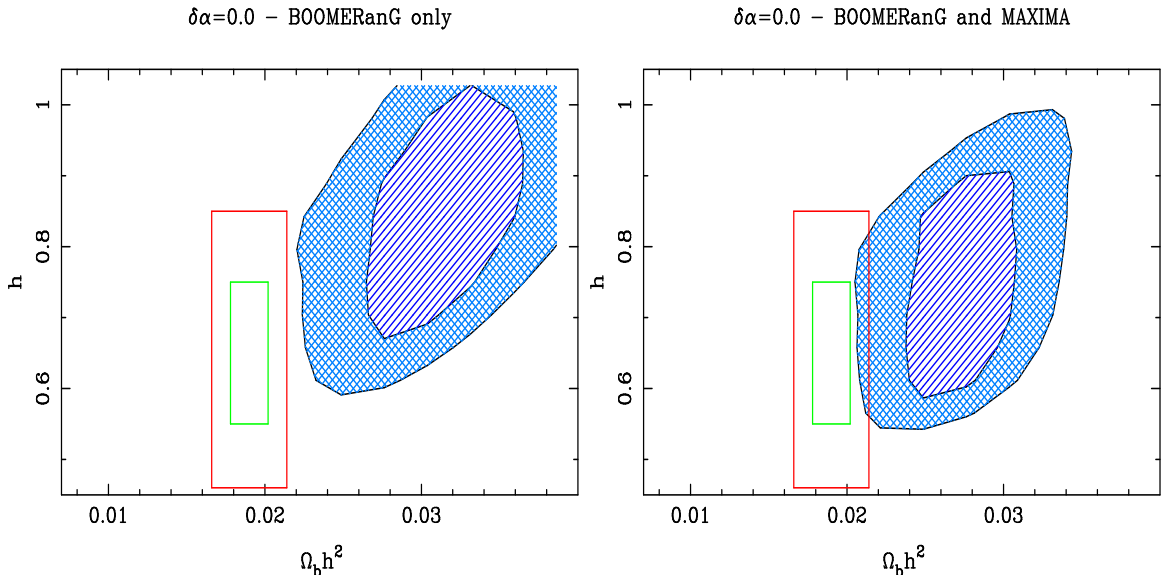

\centerline{\psfig{file=Ho_b.000.ps,angle=-90,height=3.0in,width=3.0in}
\psfig{file=Ho_b.000.B+M.ps,angle=-90,height=3.0in,width=3.0in}}
\caption{The marginalized likelihoods in the $\Omega_{\rm b}h^2$---$h$
plane for the B98 (left) and B98+M99 (right) data. Shown are the
1-$\sigma$ (68\% confidence) and 2-$\sigma$ (95\%) contours, plus a
box showing the region $h=0.65\pm 0.1$ and $\Omega_{\rm b}h^2=0.019\pm
0.0012$ and the corresponding 2-$\sigma$ contour. Note that the
regions corresponding to the direct and CMB measurements do not
overlap when one includes only B98 and there is only very little
overlap when M99 is included as well.}    
\label{fig-zeroalpha}
\end{figure}

To illustrate this point we have computed flat band power estimates
for the CMB anisotropies in the range probed by the B98 and M99
experiments for flat $\Lambda$CDM models for a grid of cosmological
parameters  ($h=0.45-1.05$, $\Delta h = 0.05$; $\Omega_{\rm
b}h^2=0.007-0.040$,   $\Delta \Omega_{\rm b}h^2= 0.003;$  $\Omega_{\rm
m}=0.2-0.8$, $\Delta \Omega_{\rm m} = 0.1;$  $n_{\rm S}=0.8-1.1$,
$\Delta n_{\rm S}=0.05$.), computed the likelihood of the models given
the data and then marginalized over the parameters $n_{\rm S}$ and
$\Omega_m$, assuming that (1) $A_{\rm S}$ be that measured by COBE
with Gaussian errors of approximately 15\%~\cite{BW}, (2) the B98
calibration errors are Gaussian, with an amplitude of $20\%$, and (3)
the M99 calibration errors have an equivalent amplitude of $8\%$ with Gaussian
correlations with respect to the other measurements when included.
The relative likelihood contours are presented in
Fig.~\ref{fig-zeroalpha} for B98 alone and for it combined with M99,
included also is a box giving an idea of where the direct limits lie.
It is clear that the CMB measurements
disagree with the direct measures of these cosmological parameters
at the $2\sigma$ level if one only takes into account B98, and this
conclusion is only slightly weakened by the inclusion of M99.  Thus, 
the CMB measurements appear to be in conflict with 
the baryon density inferred from nucleosynthesis
and either the measurements of $h$ or the
theoretical prejudice of $\Omega_{\rm tot} = 1$. 
(This is supported by the analyses of
refs.~\cite{boom1,TZ}.) 

\begin{figure}
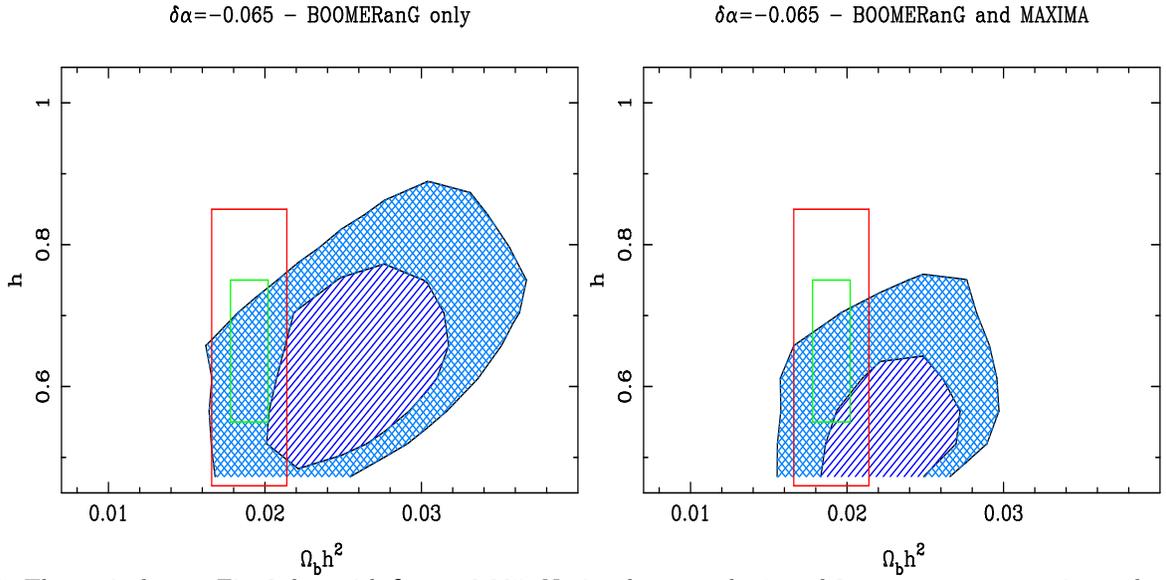

\centerline{\psfig{file=Ho_b.065m.ps,angle=-90,height=3.0in,width=3.0in}
\psfig{file=Ho_b.065m.B+M.ps,angle=-90,height=3.0in,width=3.0in}}
\caption{The equivalent to Fig.~\ref{fig-zeroalpha}, but with
$\delta\alpha=-0.065$. Notice that now the 1- and 2-$\sigma$ contours
are now in good agreement with the region preferred by direct
measurements which have been assumed to be unaffected by the change
in $\alpha$.}    
\label{fig-alpha}
\end{figure}

These conflicts can be resolved if one considers changing the 
value of $\alpha$ at last scattering.  If we repeat the above analysis, 
there is a marked improvement in the consistency of the
CMB measurements with the direct measurements of $h$ and $\Omega_{\rm
b}h^2$ when $\delta\alpha<0$. We illustrate this point in
Fig.~\ref{fig-alpha} where $\alpha$ is reduced by 6.5\% from its
present day value, that is $\delta\alpha=-0.065$. 
This shift brings the CMB into better agreement with the direct
measurements. 
It should be noted, however, that the direct measurements
themselves may also be modified by a change in the fine structure
constant, which we have not attempted to model here. This 
issue we will discuss further in section~\ref{sect-dis}.  

\begin{figure}
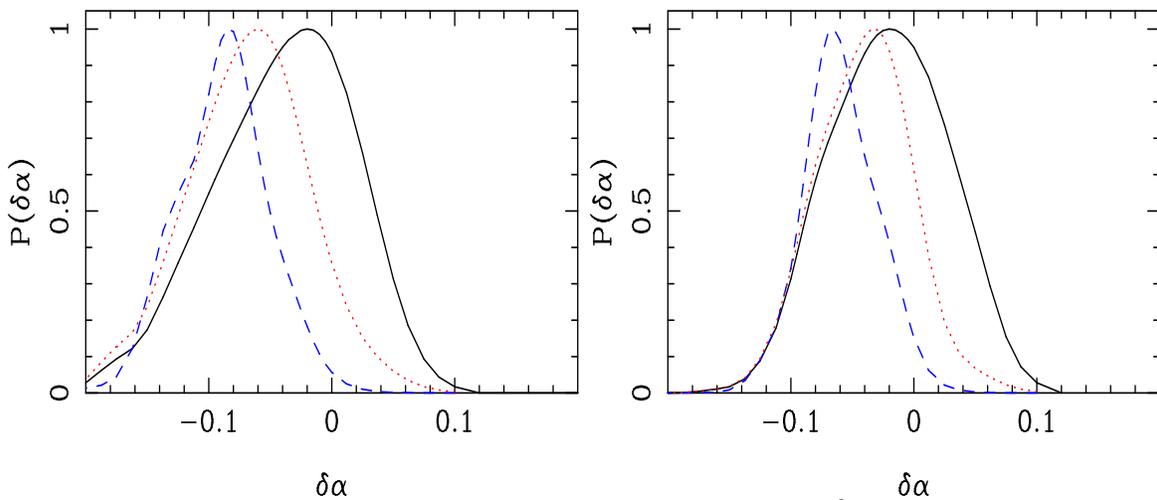

\centerline{\psfig{file=alpha.B.ps,angle=-90,height=3.0in,width=3.0in}
\psfig{file=alpha.B+M.ps,angle=-90,height=3.0in,width=3.0in}}
\caption{The likelihood marginalized over the cosmological parameters
$h$, $\Omega_{\rm m}$, $\Omega_{\rm b}h^2$, $n_{\rm S}$ and $A_{\rm
S}$ as a function of $\delta\alpha$ with no prior (P0 - solid line),
the age prior (P1 - dotted line)  and the combined $h$/strong 
BBN prior (P4 - dashed line). All the probability distributions are
biased toward $\delta\alpha<0$, with those for P4 implying a
substantial probability that $\delta\alpha<0$.}    
\label{fig-int}
\end{figure}

\begin{table}
\centering
\begin{tabular}{c|cccc|cccc}
&  \multicolumn{4}{c|}{Boomerang Only} &  \multicolumn{4}{c}{Boomerang
\& Maxima} \\ \hline Prior& $\delta \alpha$ (\%) & $68.5\%$ CL &
$95.5\%$ CL & ${\cal{P}}(\delta\alpha<0)$ & $\delta \alpha$ (\%) &
$68.5\%$ CL  & $95.5\%$ CL &${\cal{P}}(\delta\alpha<0)$ \\ \hline P0
& -2.0 &  $^{+4.5}_{-5.0}$ & $^{+8.0}_{-13}$ & $72\%$  & -2.0 &
$^{+4.5}_{-5.0}$ & $^{+8.0}_{-9.0}$ & $63\%$ \\ P1   & -6.0 &
$^{+4.0}_{-5.0}$ & $^{+8.5}_{-11}$ & $91\%$  & -3.0 &
$^{+3.0}_{-5.0}$ & $^{+7.0}_{-8.0}$ & $84\%$ \\ P2   & -5.0 &
$^{+4.0}_{-4.0}$ & $^{+7.5}_{-9.0}$ & $89\%$  & -2.5 &
$^{+2.5}_{-4.0}$ & $^{+6.0}_{-7.0}$ & $77\%$ \\ P3   & -6.5 &
$^{+4.0}_{-4.0}$ & $^{+8.0}_{-8.5}$ & $94\%$  & -3.5 &
$^{+3.5}_{-3.5}$ & $^{+6.0}_{-6.5}$ & $84\%$ \\ P4   & -8.5 &
$^{+3.0}_{-4.0}$ & $^{+7.0}_{-8.0}$ & $99\%$  & -6.5 &
$^{+3.5}_{-2.5}$ & $^{+6.5}_{-5.0}$ & $97\%$ \\ P5   & -7.0 &
$^{+3.0}_{-2.5}$ & $^{+6.0}_{-8.0}$ & $98\%$  & -5.5 &
$^{+3.0}_{-2.5}$ & $^{+5.5}_{-7.0}$ & $96\%$ \\
\end{tabular}
\caption{The peaks of the probability distribution functions for
$\delta\alpha$ for the various priors for the B98 data alone and for
the combined data sets.   Also tabulated  are the 1- and 2-$\sigma$
intervals and  the integrated probability that $\delta\alpha < 0.$
Note that all the probability distributions favor $\delta\alpha<0$.}
\label{tab-consboom}
\end{table}

To further quantify this, one can derive
likelihoods for the value of $\delta\alpha$ by marginalizing over all the
other cosmological parameters, including the Hubble constant and baryon density.
We have done this for a wide range
of $\delta\alpha$ using a slightly wider spacing than above ---
$\Delta h=0.1$ and $\Delta\Omega_{\rm b}h^2=0.006$.
In addition to the CMB data, we
consider a number of possible prior assumptions for the parameters,
particularly focusing on those involving $h$ and $\Omega_{\rm
b}h^2$. Results based on the CMB data alone, without any prior, are
labelled P0. We have included 
two simple priors incorporating fairly weak constraints on
the age of the universe  (P1: $t_0 > 11.5$ Gyr) or  on the Hubble
constant (P2: $h = 0.65 \pm 0.1$.)  These tend to have similar
effects, as both effectively cut off the  large $h$ region of
parameter space. We have also considered adding weak and strong priors on
the baryon density to the previously assumed Hubble constant prior,
(P3: P2 +  $\Omega_{\rm b}h^2 = 0.019 \pm 0.006$) and (P4: P2 +
$\Omega_{\rm b}h^2 = 0.019 \pm 0.0012$.)  Finally, we combine the age,
Hubble constant, and strong baryon density priors with a constraint
based on the cluster baryon fraction (P5: P1 + P4 +  $f_{\rm b} = 0.067 \pm
0.008$.), where $f_{\rm b}$ is the fraction of baryons to total mass deduced
from X-ray observations of rich clusters. In each case the quoted
error bars above are taken to be the $68\%$ (1-$\sigma$) confidence level.

The result of these  marginalized distributions for $\delta\alpha$ for
P0, P1 and P4 are shown in Fig.~\ref{fig-int}, and the basic
properties of these  distributions are displayed in Table I for all
the priors.  As can be seen, in the absence of any priors the data
prefer a value for  $\alpha$ at last scattering a few percent lower
than its present value, but the constraint is fairly weak.
Considering only the B99 data and adding
fairly weak constraints on the age or $h$ gives a
significantly stronger signal, suggesting a detection of variation in
$\alpha$  at the 1-$\sigma$ level.  Finally, if one includes the
stronger constraints that the baryon density is low, then the
evidence for variation in $\alpha$ is significant at the  2-$\sigma$
level.  Including the M99 data weakens these detections somewhat, but
not dramatically.  

One can understand the effects of the priors in
light of our earlier discussion of how the CMB
anisotropies depend on $\delta\alpha$. By allowing $h$ and
$\Omega_{\rm b}h^2$ to vary freely one can fit both the primary
peak position and secondary peak heights either 
with $\delta\alpha=0$ or $\delta\alpha\ne 0$, and 
as we have already seen in Fig.~\ref{fig-zeroalpha},
fixing $\delta\alpha=0$ requires large values of $h$ and
$\Omega_{\rm b}h^2$. The constraint on $h$ comes primarily from the
position of the primary peak.  When  
one allows the time of recombination to change due to a variation in
$\delta\alpha$, the high value of $h$ can be offset by an increase in
$c_{\rm s}\eta_*$, making models with a lower values of $h$ and
$\delta\alpha<0$ equally likely. If one further assumes a prior 
such as P1 or P2 which penalizes high $h$, the models with
$\delta\alpha<0$ become favored over those with $\delta\alpha=0$. 
A similar line of argument can be applied to $\Omega_{\rm b}h^2$. 
When $\delta\alpha=0$ a high value of $\Omega_{\rm
b}h^2(\approx 0.031)$ is favored, but when one allows $\delta\alpha$
to vary, the approximate degeneracy between $\delta\alpha$ and
$\Omega_{\rm b}h^2$ makes models with a lower value of $\Omega_{\rm
b}h^2$ and $\delta\alpha<0$ equally likely. Clearly, the
inclusion of a constraint which penalizes high $\Omega_{\rm b}h^2$
will favor models with $\delta\alpha<0$. The remarkable aspect of this
is that the inclusion of a single parameter, $\delta\alpha$, has the
effect of improving the fit to the observations through two physically
very distinct mechanisms.

\begin{figure}
\centerline{\psfig{file=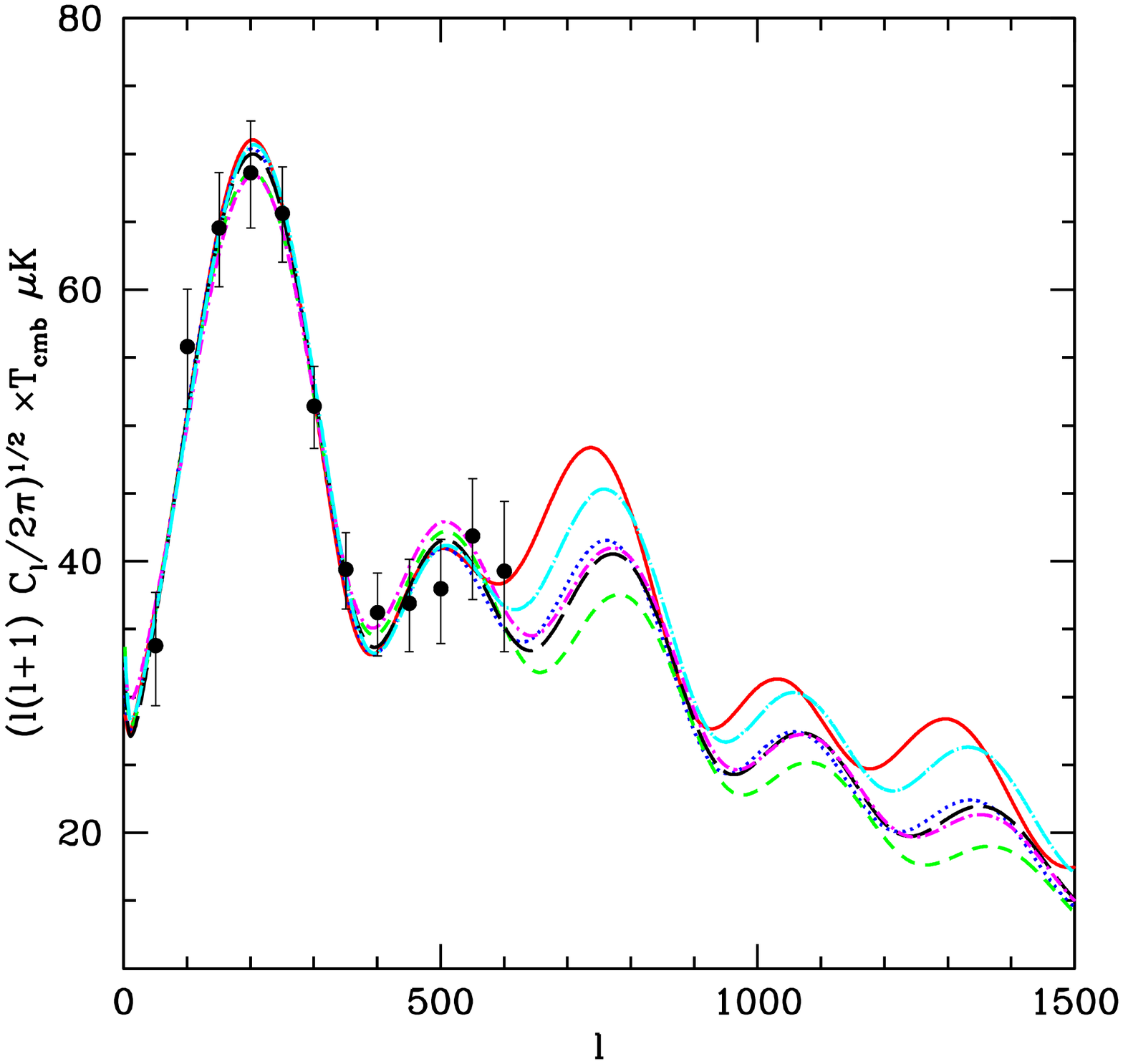,height=3.0in,width=3.0in}
\psfig{file=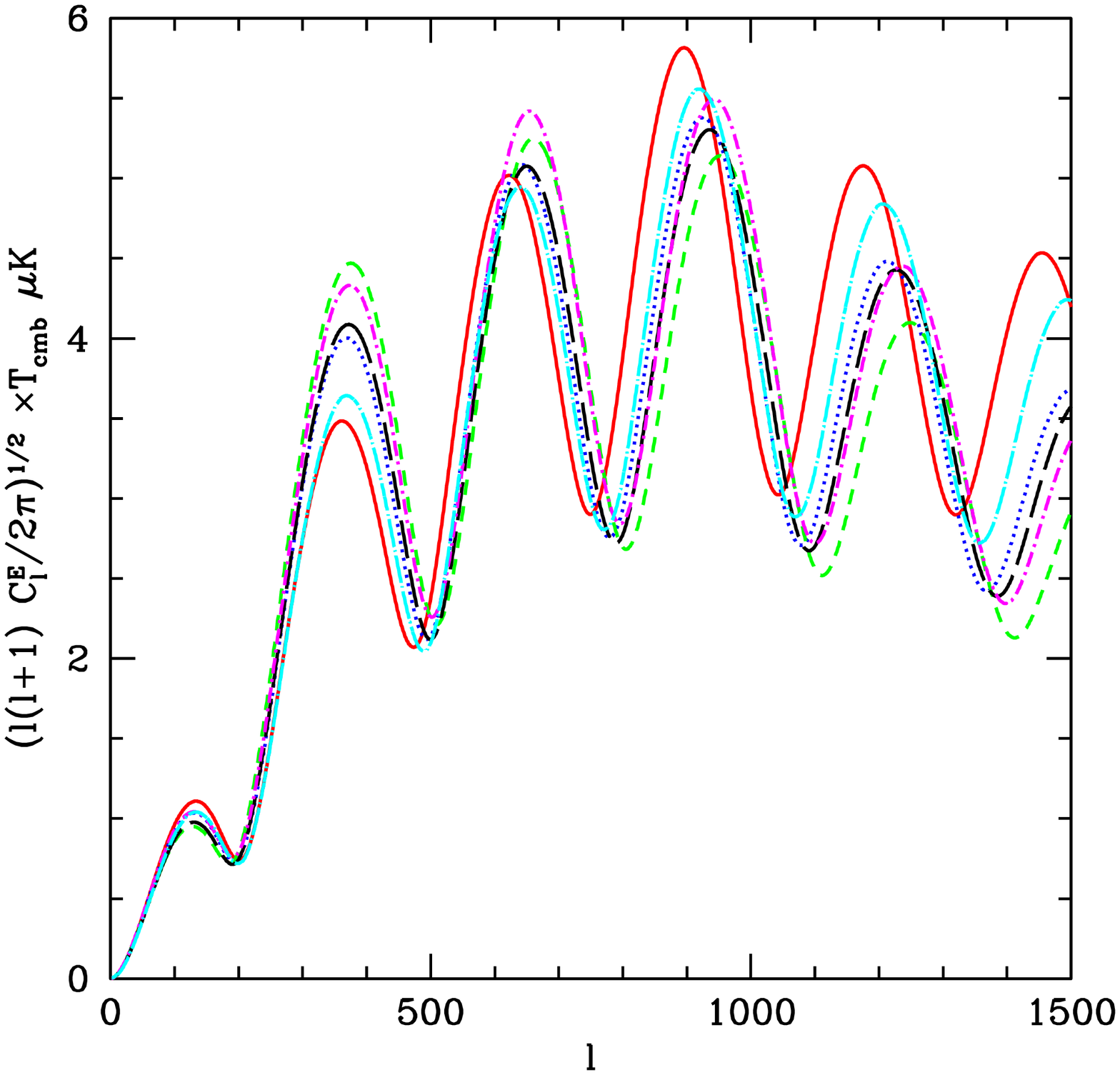,height=3.0in,width=3.0in}}
\caption{The angular power spectra of temperature anisotropies and
polarization for the best fit models to the B98 data from Table
II. Included are P0 with $\delta\alpha=0$ (Long dashed-dot line) and P1
(Solid), P2 (dotted), P3/P4 (short dashed) P5 (short-dash dotted) all
with $\delta\alpha\ne 0$, along with the B98 data. The best fit
normalization changes for different models, therefore each curve has
been appropriately normalized so that they can all be plotted with the
data on the same graph.}    
\label{fig-best-b98}
\end{figure}

\begin{figure}
\centerline{\psfig{file=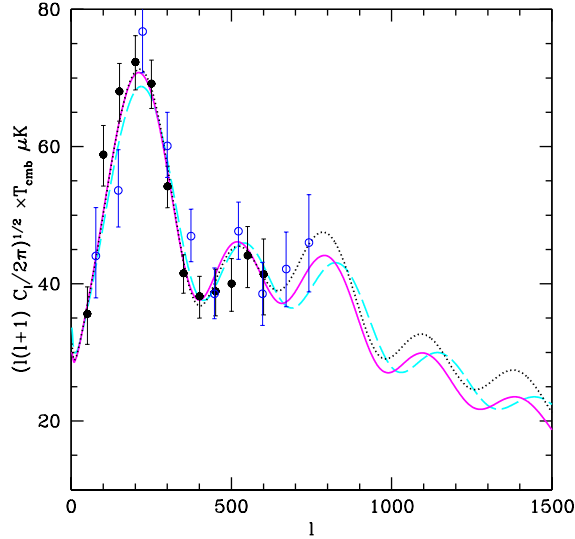,height=3.0in,width=3.0in}}
\caption{The angular power spectra of temperature anisotropies 
for the best fit models to the B98 and M99 data from Table
III. Included are P0 with $\delta\alpha=0$ (dotted line) and P4 with 
$\delta\alpha=0$ (dashed line) with $\delta\alpha\ne 0$ (solid line). 
With the strong BBN constraint on the baryon density, the 
$\delta\alpha\ne 0$ case provides a much better fit than the 
$\delta\alpha=0$ case. As for Fig.~\ref{fig-best-b98} we have
normalized so that all the models can be presented on the
same plot. In doing this we have fixed the calibration of the
B98 data relative to the M99 calibration to be 15\% higher than its 
nominal value.}
\label{fig-best-bm}
\end{figure}

The conclusions are substantiated by examining the best fit models
listed in Tables II and III, and plotted compared to the B98 data 
in Fig.\ref{fig-best-b98} and to the combined dataset in
Fig.~\ref{fig-best-bm}. Except for P0 in the case of B98 alone, and
P0 and P1 in the case of B98 and M99, the reduced $\chi^2$ of the 
best fits are decreased by allowing $\alpha$ to vary, but the fits are
only significantly improved  when one assumes a strong prior for the
baryon density (P3-P5). When considering just the B98 data, the $\chi^2$
appears to be somewhat smaller than the number of degrees of freedom,
so that the reduced $\chi^2$ are significantly less than one, which
in turn suggests that the error bars of the B98 data may be overestimated. 
The reduced  $\chi^2$ no longer appear low when the M99 data are 
included in the analysis.

\begin{table}
\centering
\begin{tabular}{cccccccc}
\hline Prior& $\delta\alpha$ & $h$ & $\Omega_{\rm b}h^2$ &
$\Omega_{\rm m}$ & $n_{\rm S}$ & $R_{\rm B98}$ & $\chi^2$ \\ \hline P0
& 0.0 & 0.95 & 0.031 & 0.2 & 0.925 & 1.00 & 4.03\\ & -0.020  & 0.85 &
0.031 & 0.3 & 0.975 & 0.92 & 3.90 \\ P1 & 0.0 & 0.85 & 0.025 & 0.2 &
0.850 & 1.12 & 5.09\\ & -0.055 & 0.75 & 0.025 & 0.2 & 0.900 & 0.92 &
4.00 \\ P2& 0.0 & 0.75 & 0.031 & 0.6 & 0.975 & 1.00  & 5.66 \\ &
-0.070 & 0.65 & 0.025 & 0.3 & 0.900 & 0.94  & 4.10 \\ P3& 0.0 & 0.65 &
0.025 & 0.6 & 0.900 & 1.14  & 7.84\\ & -0.080 & 0.65 & 0.019 & 0.2 &
0.850 & 0.98  & 4.29\\ P4& 0.0 & 0.75 & 0.019 & 0.2 & 0.800 & 1.16  &
10.15 \\ & -0.080 & 0.65 & 0.019 & 0.2 & 0.850 & 0.98  & 4.29 \\ P5&
0.0 & 0.65 & 0.019 & 0.4 & 0.800 & 1.32 & 12.13 \\ & -0.070 & 0.55 &
0.019 & 0.4 & 0.850 & 1.08 & 6.60 \\ \hline
\end{tabular}
\caption{The best fit models for the B98 data only with the various
priors, with and without allowing variations in $\alpha$. The number
of degrees of freedom (number of data points minus number of theoretical 
parameters) for the fits are roughly between 7 and 10 when $\delta\alpha=0$, 
depending on the number of constraints.  (This is reduced by 1 for the 
$\delta\alpha \ne 0$ models.)  Models with stronger priors included in the 
likelihood
effectively have more data and thus more degrees of freedom.  
Note that the fits are substantially improved by
allowing $\delta\alpha\ne 0$ when there is a prior assumption on
$\Omega_{\rm b}h^2$ and that the reduced $\chi^2$ are
generally less than one. $R_{\rm B98}$ is the ratio of the COBE and
B98 normalizations for the $C_{\ell}$'s.}
\label{tab-bestboom}
\end{table}

\begin{table}
\centering
\begin{tabular}{ccccccccc}
\hline Prior& $\delta\alpha$ & $h$ & $\Omega_{\rm b}h^2$ &
$\Omega_{\rm m}$ & $n_{\rm S}$ & $R_{\rm B98}$  & $R_{\rm M99}$ &
$\chi^2$\\ \hline P0 & 0.0 & 0.75 & 0.025 & 0.3 & 0.925 & 0.92 & 1.06
& 16.43 \\ & -0.025  & 0.65 & 0.025 & 0.4 & 0.925 & 0.92 & 1.06 &
15.82\\ P1& 0.0 & 0.75 & 0.025 & 0.3 & 0.925 & 0.92 & 1.06 & 16.43\\ &
-0.025 & 0.65 & 0.025 & 0.4 & 0.925 & 0.92 & 1.06 & 15.82\\ P2& 0.0 &
0.65 & 0.025 & 0.5 & 0.925 & 0.98 & 1.09 & 17.13\\ & -0.025 & 0.65 &
0.025 & 0.4 & 0.925 & 0.92 & 1.06 & 15.82\\ P3& 0.0 & 0.65 & 0.025 &
0.5 & 0.925 & 0.98 & 1.09 & 18.13\\ & -0.025 & 0.65 & 0.025 & 0.4 &
0.925 & 0.92 & 1.06 & 16.82\\ P4& 0.0 & 0.75 & 0.019 & 0.2 & 0.850 &
1.00 & 1.15 & 21.89\\ & -0.065 & 0.65 & 0.019 & 0.2 & 0.900 & 0.82 &
0.97 & 18.08\\ P5& 0.0 & 0.65 & 0.019 & 0.3 & 0.850 & 1.00 & 1.15 &
24.43\\ & -0.055 & 0.55 & 0.019 & 0.4 & 0.900 & 0.90 & 1.03 & 18.96\\
\hline
\end{tabular}
\caption{The best fit models for the B98 and M89 data with the various
priors, with $\delta\alpha=0$ and $\delta\alpha\ne 0$.  The number of
degrees of freedom for the fits are in the range 16-19 for  
$\delta\alpha=0$. (Again, it is reduced by 1 when $\delta\alpha\ne 0$). 
$R_{\rm M99}$ is the ratio of the COBE and M99
normalizations for the $C_{\ell}$s.}
\label{tab-bestmax}
\end{table}

\section {Discussion}
\label{sect-dis}

In the previous section we presented evidence for a variation in
$\alpha$ using recent CMB observations. If confirmed by subsequent
observations this would be a truly remarkable 
result. In this section we will discuss the various aspects of our
analysis, focusing on the potential uncertainties. 

First, we should make some comments on the details of our
statistical procedure and the models which we have probed.  We make
the approximation that the CMB data points are statistically
independent and Gaussianly distributed, with window functions given by
flat band powers in $\ell$.  This is also an assumption in the
analysis of ref.~\cite{TZ}, but not in those of the BOOMERanG and
MAXIMA collaborations~\cite{boom2,maxima2} where the exact
experimental window functions were used. 
Since the analysis presented here for $\delta\alpha=0$  
agrees qualitatively with those other analyses, we believe 
these approximations should be sufficient for our
purposes.  The B98 data has more points (12 versus 10) and smaller
errors than the  M99 data, and generally provides stronger
constraints.   While both data sets suggest the secondary peak is
suppressed relative to the  first Doppler peak, the M99 data shows no
evidence for a left-ward shift of the peak.  Thus, including it tends
to weaken the evidence for a  time varying $\alpha.$  

We include in our calculations the uncertainties in the absolute
calibrations of the data sets, which is necessary in order for them to
be consistent with each other.  The B98 data was normalized by the CMB
dipole, which is subject to large systematic errors, and their quoted
calibration error is 20\% in the power.  The M99 experiment was also
able calibrate off of Jupiter, and has only an 8\% error.  Best fit
models allowing both of these to vary seem to prefer an increase of
roughly 15\% in the relative B98/M99 power calibration (see, for
example, Table III).  

We should also note that the range of models we use in our analysis
does not include many cosmological scenarios that are often
considered.  We have excluded the possibility of tensor fluctuations
with a spectral index $n_{\rm T}$ and amplitude $A_{\rm T}$, a hot
dark matter component $\Omega_{\nu}$, and also that  of early
reionization often quantified by $\tau_{\rm R}$, the optical depth to
reionization. These were included in ref.\cite{TZ} and were found to
have little or no bearing on the preferred values of $\Omega_{\rm
b}h^2$ and $h$ since these parameters are effectively orthogonal 
given the present data.  
In fact, in ref.~\cite{boom2} it was suggested that
there is a degeneracy between $n_{\rm S}$ and $\tau_{\rm R}$ for the
angular scales probed in B98; our unusually low values of $n_{\rm S}$,
therefore, take into account the possibility of reionization at moderate
redshift with $n_{\rm S}=1$ being compatible with the data. 

We have investigated this by constructing the Fisher matrix which
quantifies the effects of changing parameters has on the measured band
powers ${\cal B}_i$, 
\begin{equation} 
{\cal{F}}_{ab} = {\partial {\cal B}_{i} \over \partial p_a} C^{-1}_{ij} 
{\partial {\cal B}_{j} \over \partial p_b},  
\end{equation}  
where $p_a$ are the parameters ($\Omega_{\rm m}h^2, \Omega_{\rm b}h^2,
h, \delta\alpha, n_{\rm S}, \tau_{\rm R}, A_S$) and $C_{ij}$ is the
data  covariance matrix, assumed diagonal except for the calibration
uncertainties.  We find the strongest degeneracy of $\delta\alpha$ to
be with  $\Omega_{\rm m}h^2$, but there are also significant overlaps
with  $h$ and $\Omega_{\rm b}h^2$; all of which are consistent with 
the simple theoretical arguments in Section~\ref{sect-cmb}.  Our
inferred errors on $\delta\alpha$ from  Figure 4 are quite consistent
with those expected by computing inverse Fisher  matrix for the best
fit model with P0. The matrix is also largely block diagonal as one
might have expected, with  $n_{\rm S}, \tau_{\rm R}, A_S$ being
largely orthogonal to the other variables, but with significant
overlap amongst themselves, confirming that  $n_{\rm S}$ and
$\tau_{\rm R}$ are degenerate given the present data.  

Knowledge of the Fisher matrix  allows us to investigate the impact
future  CMB measurements might have on further constraining
parameters.  If the  error bars of the B98 and M99 experiments are 
reduced by a factor of two, the errors on $\alpha$ (and indeed 
on most parameters) are reduced by a comparable factor.
Hence, improved sensitivity with the same angular coverage is an
important goal.  We have also considered the impact of a hypothetical
measurement of the third peak, centered at $\ell = 800$, as well as a
detection of the first polarization peak centered at $\ell = 350$,
assuming 10\% errors on a flat band power measurement in each case.  Both,
particularly the polarization measurement, help to reduce
uncertainties in $\Omega_{\rm m}h^2$ and $\Omega_{\rm b}h^2$,   but
disappointingly neither do very well in reducing the uncertainties in
$\delta\alpha$. It should be noted that a weakness of this band power
based approach is that the  conclusions will depend somewhat on how
the data are binned, particularly  if the models vary greatly across
the bins.
     
Having argued that our statistical procedure and the models which we
have probed provide a robust detection of a variation in $\alpha$
given prior assumptions from direct measurements of $h$ and
$\Omega_{\rm b}h^2$ in a flat universe, we now turn to the more
difficult issue of the effect of such a variation on these direct
measurements. This pertains primarily to those associated  with BBN
since the measurements of $h$, $t_0$ and $f_b$ are made at very
low redshifts and therefore will be relatively insensitive to these
changes.

Since electromagnetic effects are ubiquitous in BBN, it is clear that
there must exist a constraint on $\delta\alpha(t_{\rm nuc})$ from the
consistency of BBN with light element abundances. There are two
approaches to this problem documented in the literature.  The
first~\cite{KPW,Barrow} is to use only the observations of $^4{\rm
He}$, which are thought to be the most reliable.
One can make a very simple estimate of the
primordial $^4{\rm He}$ abundance, in terms of $m_n/m_p$, the neutron
to proton mass ratio. However, expressing this in terms of $\alpha$
cannot be done in a model independent way because it involves a subtle
interplay between electromagnetic, weak and strong interaction effects which
have not been understood completely within QCD. Therefore, tight
limits on $\delta\alpha(t_{\rm nuc})$ computed in this way should be
treated with some caution. 

A more reliable alternative~\cite{Berg} is
to make a detailed analysis of how changes in $\alpha$ can effect
all the light abundances and derive a constraint from demanding
consistency with their observed values.  Although
this involves a number of complicated nuclear reaction rates, 
it turns out that a model
independent constraint might be possible at the level of around a few
percent. In fact, it was suggested in ref.~\cite{Berg} that
$|\delta\alpha(t_{\rm nuc})|\approx 0.02$ could help BBN fit the
observed light element abundances better. 
However, even this value should be treated with
some caution since there are further uncertainties which might could 
modify it by as much as a factor of two~\cite{Lopez}.

Furthermore, the implied value of $\Omega_{\rm b}h^2$ is likely to
change as a function of $\delta\alpha(t_{\rm nuc})$. For example, 
one might expect
that~\cite{Lopez}
\begin{equation}
\Omega_{\rm b}h^2\left[\delta\alpha(t_{\rm
nuc})\right]=\left[1+A\delta\alpha(t_{\rm nuc})\right]\Omega_{\rm
b}h^2\left[0\right]
\end{equation}
for small values of $\delta\alpha(t_{\rm nuc})$ where 
the coefficient is expected to be of order $A\sim{\cal
O}(1)$. 
The amplitude and sign of the proportionality constant $A$ will 
clearly have some influence on our conclusions.
In particular, if decreasing $\alpha$ increases the inferred baryon
density $(A < 0)$, then it may be possible to fit the data with a
smaller change in $\alpha$.  However, if the opposite is true $(A >
0)$, then it may prove a better fit if the fine structure constant is
larger at last scattering, contrary to our present results. 

To make contact with the earlier discussion, one needs  to relate
$\delta\alpha(t_{\rm nuc})$ and $\delta\alpha(t_{\rm rec})$ which
requires a model for how the variation in $\alpha$ is realized.  If
$\alpha$ is increasing with time as we have suggested here, then it
might be sensible to assume that it has done so
monotonically\footnote{In fact models have been suggested in which
$\alpha$ oscillates~\cite{oscillate}, although this would appear at
this stage to be somewhat {\it ad hoc}.} and, therefore,
$\delta\alpha(t_{\rm nuc})<\delta\alpha(t_{\rm rec})$.  Given the
uncertainties, the constraint from BBN on $\delta\alpha(t_{\rm nuc})$
would appear to  be consistent with this relation given the fairly
small  values of $\delta\alpha(t_{\rm rec})$ required for a good fit
to the CMB data.  However, this certainly motivates a critical
appraisal of the exact constraint on $\delta\alpha$ from BBN.
Including the effect of changing $\alpha$ on the BBN measurements
would require knowledge of the  parameter $A$ and we have not
attempted to incorporate this into our analysis. This issue should be
revisited in future work when the  $\alpha$ dependence of the BBN
constraints are better understood.

Finally, we should mention that realistic models in which $\alpha$ varies may
contain one or more light scalar fields which mediate the precise
variation. Clearly, if such a field exists it should be included in
the calculation of the CMB anisotropies in the Boltzmann hierarchy of
CMBFAST, either explicitly or as a extra relativistic degree of
freedom. This would allow a subsequent analysis to include effects of
the time variation of $\alpha$, rather than just a change between the
time of recombination and the present day.   Such a field could have
significant  energy density during the epochs important for structure
formation (after the time of radiation-matter equality) and it  may
even be possible for such a field to act as a quintessence
field~\cite{MB}, removing the need for $\Lambda$.

\section{Conclusions}
\label{sect-con}

The most recent CMB data provide strong evidence yet that the universe
is, at least approximately, spatially flat.  The B98 data, however, is
not entirely  consistent with spatial flatness and direct measurements
of other  cosmological parameters.  The situation is only slightly
improved when the  M99 data is included.  However, if the fine
structure constant was a  few percent smaller when the photons were
last scattered, then a model can  be found which is consistent with
all observations. It is clear that a change in the  $\alpha$ is not
the only possible explanation for such observations, and the evidence
we present here could equally well be thought of as favoring other
delayed recombination models, for example, that presented in
ref.~\cite{PSH}.

The evidence for a time variation in the fine structure constant is
significant when a tight prior is assumed for the baryon density.
However, the baryon density inferred from measurements of primordial
abundances depends on nuclear physics processes at times long before
last scattering at the epoch of BBN. We have argued that the values of
$\delta \alpha(t_{\rm rec}) \sim -0.05$  that we have deduced are
consistent with BBN  given the uncertainties assuming that the
variation in $\alpha$ is monotonic, but that inclusion of the effect
of varying $\alpha$ on the inferred value of $\Omega_{\rm b}h^2$ for a
given Deuterium abundance has been ignored, mainly due to lack of
quantitative information.

Stronger conclusions must, of course,  wait for better data, such as
might come from the  satellite experiments MAP and PLANCK.  In
particular, these will be able  to confirm whether the inconsistencies
of flat models with direct measurements  and the CMB data (such as a
slight shift of features to larger scales)  are real.  In addition,
these experiments should be able to break the  degeneracy between a
changing $\alpha$ and $\Omega_{\rm b}h^2$, so that  a change in
$\alpha$ can be tested independently of what occurred at
nucleosynthesis. It is clear from Fig.~\ref{fig-best-b98} that the
best fit models can differ greatly for $\ell>600$ in temperature
anisotropies and the polarization; any experiment which  probes the
CMB in these areas will be useful in breaking these degeneracies,
although the results of our Fisher matrix analysis suggest they will
require considerable sensitivity. 

Having presented our case for a few percent variation in $\alpha$ at
around $z\approx 1000$, it is interesting to compare to the other
claimed detection of a change in $\alpha$ around $z\approx 1$, and
speculate as to an explanation. Naively, $\delta\alpha(z\approx
1000)\sim  - 0.01$ and $\delta\alpha(z\approx 1)\sim - 10^{-5}$ might
suggest that $\delta\alpha\propto z$, but this would be 
incompatible with BBN if extrapolated back to $z\approx
10^9-10^{10}$. In order to have a chance of being consistent with BBN,
such a scenario would require that the variation in $\alpha$
terminated at some point shortly before recombination, that is,
$\delta\alpha(t_{\rm nuc})\sim\delta\alpha(t_{\rm rec})$. At first
sight coupling $\alpha$ non-minimally to gravity via the Ricci scalar
or the trace of the energy-momentum tensor as suggested in
ref.~\cite{Beck}, both of which are zero in the radiation-era and
non-zero in the matter-era, might seem an attractive solution since
the variation in $\alpha$ would begin at the onset of matter
domination. Clearly, such ideas could have a profound impact on
understanding of Grand Unification and this particular  interpretation
of the most recent observations presented here opens up a wide range
of interesting possibilities for future research in this area.

\acknowledgements  We wish to thank J. Barrow, G. Efstathiou, N. Turok
and particularly R. Lopez for useful  conversations. RB and RC were
supported by PPARC Advanced Fellowships, while JW is supported by the
German Academic Exchange Service (DAAD), DOE grant DE-FG03-91ER40674
and UC Davis. The computations were performed on COSMOS, at the UK
National Cosmology Computing Centre in Cambridge.  
 
\def\jnl#1#2#3#4#5#6{\hang{#1, {\it #4\/} {\bf #5}, #6 (#2).} }
\def\jnltwo#1#2#3#4#5#6#7#8{\hang{#1, {\it #4\/} {\bf #5}, #6; {\it
ibid} {\bf #7} #8 (#2).} } \def\prep#1#2#3#4{\hang{#1, #4.} }
\def\proc#1#2#3#4#5#6{{#1 [#2], in {\it #4\/}, #5, eds.\ (#6).} }
\def\book#1#2#3#4{\hang{#1, {\it #3\/} (#4, #2).} }
\def\jnlerr#1#2#3#4#5#6#7#8{\hang{#1 [#2], {\it #4\/} {\bf #5}, #6.
{Erratum:} {\it #4\/} {\bf #7}, #8.} } \def\prl{Phys.\ Rev.\ Lett.}
\def\pr{Phys.\ Rev.}  \def\pl{Phys.\ Lett.}  \def\np{Nucl.\ Phys.}
\def\prp{Phys.\ Rep.}  \def\rmp{Rev.\ Mod.\ Phys.}  \def\cmp{Comm.\
Math.\ Phys.}  \def\mpl{Mod.\ Phys.\ Lett.}  \def\apj{Astrophys.\ J.}
\def\apjl{Ap.\ J.\ Lett.}  \def\aap{Astron.\ Ap.}  \def\cqg{Class.\
Quant.\ Grav.}  \def\grg{Gen.\ Rel.\ Grav.}  \def\mn{Mon.\ Not.\ Roy.\
Astro.\ Soc.}
\def\ptp{Prog.\ Theor.\ Phys.}  \def\jetp{Sov.\ Phys.\ JETP}
\def\jetpl{JETP Lett.}  \def\jmp{J.\ Math.\ Phys.}  \def\zpc{Z.\
Phys.\ C} \def\cupress{Cambridge University Press} \def\pup{Princeton
University Press} \def\wss{World Scientific, Singapore}
\def\oup{Oxford University Press}

\pagebreak
\pagestyle{empty}

\end{document}